\documentclass{jpp}
\usepackage{graphicx}

\usepackage[utf8]{inputenc}
\usepackage[T1]{fontenc}
\usepackage{amsmath}

\newcommand{\beq}{\begin{equation}}
\newcommand{\eeq}{\end{equation}}

\newcommand{\vu}{\boldsymbol{u}}
\newcommand{\vB}{\boldsymbol{B}}
\newcommand{\vb}{\boldsymbol{b}}

\newcommand{\vAy}{v_{\mathrm{A}y}}

\newcommand{\vz}{\boldsymbol{z}}

\newcommand{\din}{\delta_{in}}
\DeclareMathOperator{\sech}{sech}

\shorttitle{Effect of shear flow on the resistive tearing instability}
\shortauthor{A. Mallet, S. Eriksson, M. Swisdak, J. Juno}

\title{The effect of shear flow on the resistive tearing instability}

\author{Alfred Mallet\aff{1}
  \corresp{\email{alfred.mallet@berkeley.edu}},
  Stefan Eriksson\aff{2},
 Marc Swisdak\aff{3},
 James Juno\aff{4}}

\affiliation{\aff{1}Space Sciences Laboratory, University of California, Berkeley CA 94720, USA
\aff{2}Laboratory for Atmospheric and Space Physics, University of Colorado, Boulder, CO 80303,USA
\aff{3}Institute for Research in Electronics and Applied Physics, University of Maryland, College Park, MD 20742, USA
\aff{4}Princeton Plasma Physics Laboratory, Princeton, NJ 08543, USA}
\begin{document}

\maketitle
\begin{abstract}
We develop a new scaling theory for the resistive tearing mode instability of a current sheet with a strong shear flow across the layer. The growth rate decreases with increasing flow shear and is completely stabilized as the shear flow becomes Alfv\'enic: both in the constant-$\Psi$ regime, as in previous results, but we also show that the growth rate is in fact suppressed more strongly in the nonconstant-$\Psi$ regime. As a consequence, for sufficiently large flow shear, the maximum of the growth rate is always affected by the shear suppression, and the wavenumber at which this maximum growth rate is attained is an increasing function of the strength of the flow shear. These results may be important for the onset of reconnection in imbalanced MHD turbulence.
\end{abstract}

\section{Introduction}

Current sheets (CS) are characterized by sharp rotation of the magnetic field across a narrow CS width $a$; often, they are extended in their other dimensions, with a CS length $L\gg a$. They form generically in many plasmas \citep{chapman1963}, and have more recently been shown to form naturally at the small scales of magnetohydrodynamic (MHD) turbulence \citep{boldyrev,mallet3d,ms16}, the motivation for this paper. Within the framework of resistive MHD, the CS width $a$ is regularized by the resistivity $\eta$: below a certain scale, the magnetic field is no longer frozen into the flow, leading to magnetic reconnection. %Originally, it was suggested by \cite{sweet1958} and \cite{parker} that this leads to a (slow) characteristic reconnection timescale proportional to $\tau_{SP} \sim (L/\vA) S^{-1/2}$, where $\vA = B/\sqrt{4\pi\rho}$ is the Alfv\'en velocity and $S=\vA L / \eta$ is the Lundquist number. 
Reconnection is important in a wide range of physical situations, for example sawtooth crashes in tokamaks \citep{kadomtsev1975} and solar flares \citep{yan2022}. Typically, the Sweet-Parker theory \citep{sweet1958,parker1957} of current sheet reconnection predicts a reconnection rate orders of magnitude too slow to explain the observed fast reconnection.

However, CS are usually unstable to the tearing instability \citep{fkr,coppi}. It has been realised that this instability means that CS cannot thin below the width $a_{crit}$ at which the tearing growth rate becomes comparable to the sheet formation timescale, resulting in fast "plasmoid-mediated" reconnection \citep{pucci2014,uzdensky2016,comisso2017}. This theory has been applied to the current sheets formed dynamically in MHD turbulence, resulting in the prediction of a dramatic change in the turbulent cascade below a certain scale \citep{msc_disruption,loureiroboldyrev}. This was recently confirmed in high-resolution MHD turbulence simulations \citep{dong2022}. 

The standard theory of the tearing mode involves a magnetic equilibrium parametrized by
\begin{equation}
b_{0y}(x) = \hat{\boldsymbol{y}}\vAy f(x/a),\label{eq:eqb}
\end{equation}
where we have written the magnetic field in velocity units $\vb = \vB/\sqrt{4\pi n_{i} m_i}$. Additionally, it is usually assumed that there is no equilibrium flow, $u_0=0$. This results in the growth rates
\begin{align}
\frac{\gamma a}{\vAy}\sim \begin{cases}
(ka)^{2/5} (\Delta'a)^{4/5}S^{-3/5}, &\Delta'\delta_{in}\ll 1 \\
(ka)^{2/3}S^{-1/3}, &\Delta'\delta_{in}\gg 1,\label{eq:resnoflow}
\end{cases}
\end{align}
where $S\equiv a \vAy/\eta$ is the Lundquist number and $\eta$ is the resistivity. We assume $S\gg1$, so that the width of the inner layer where resistivity becomes important $\delta_{in}\ll a$. $\Delta'$ is a measure of the ideal discontinuity regularized by the resistivity within the inner layer, and $\Delta'>0$ is required for instability. For sufficiently long-wavelength modes, 
\begin{equation}
    \Delta'a \propto (k a)^{-n},\label{eq:dprimen}
\end{equation}
with $n=1$ for a Harris type equilibrium with $f(x/a) = \tanh(x/a)$ and $n=2$ for $f(x/a) = \sin(x/a)$. For $\Delta'\delta_{in}\ll1$, the growth rate $\gamma$ decreases with $k$, while at $\Delta'\delta_{in}\gg1$, $\gamma$ increases with $k$. The maximum growth rate occurs where these two expressions meet: choosing $n=1$ in (\ref{eq:dprimen}),
\begin{equation}
    \frac{\gamma_{max}a}{\vAy}\sim S^{-1/2}, \quad k_{max}a\sim S^{-1/4}.\label{eq:maxnoshear}
\end{equation}
%For this mode to be accessible, $k_{max}$ must fit into the sheet,
%\begin{equation}
%    \frac{L}{a} \gtrsim \frac{1}{k_{max} a} \sim S^{1/4}.
%\end{equation}

In many situations, including the aforementioned case of turbulence, the assumption that $u_0=0$ is not realistic. Since the magnetic field is frozen into the plasma flow far from the narrow resistive layer, a shear flow across the current sheet often has a dramatic effect on the instability. This situation is the subject of the present paper. Following \cite{boldyrev2018}, we assume for simplicity that the shear flow profile is proportional to the magnetic field given in Eq.~(\ref{eq:eqb}):
\begin{equation}
    u_{0y} = \alpha b_{0y}.\label{eq:equ}
\end{equation} 
For the tearing mode, we are interested in the range $0<|\alpha|<1$: for $|\alpha|\geq1$, the tearing mode is stable and we have instead the ideal Kelvin-Helmholtz instability, with a large growth rate $\gamma_{KH}\sim (1-\alpha^2)^{1/2}\alpha\vAy/a$. Previous analytic work \citep{hofman1975,chen1989,boldyrev2018} incorporating flow shear has been largely focused on the weakly unstable $\Delta'\din \ll 1$ case, and found that the shear flow suppresses the instability and alters the scaling of the growth rate with resistivity. Specifically, \cite{chen1989} found that
\begin{equation}
\frac{\gamma a}{\vAy} = \frac{1}{\sqrt{2\pi}} \Delta'^{1/2}[\alpha(1-\alpha^2)]^{1/2}S^{-1/2},\quad  \Delta'\delta_{in} \ll 1.
\end{equation}
How shear flow affects the opposite limit, when $\Delta'\din\gg1$, is still unknown. It was recently argued by \cite{boldyrev2018} that for $\Delta'\din\gg1$ the shear flow cannot affect the growth rate, which as a consequence would mean that the maximum of the growth rate is unaffected by shear flow. Here, we will find with both scaling arguments and numerical simulations that there is a shear-modified regime for $\Delta'\din\gg1$, provided that $\alpha$ is larger than a critical value, and thus that the maximum growth rate can also be suppressed by shear flow.

Finally, we note that we have recently also analysed the collisionless tearing instability with flow shear \citep{mallet2025}, which ironically turns out to be easier to attack analytically. Because of the separation between the ion and electron scales in the collisionless problem, we were able to find an analytic solution. Irrespective of $\Delta'\din$, the growth rate is slow compared to the shear across the ion layer, but fast compared to the shear across the electron layer, and matching the solutions in both regions gives the numerically-observed growth rates. In the resistive case studied here, the growth rate  is only small compared to the shear across the resistive inner layer of width $\din$ for $\Delta'\din\ll1$. For $\Delta'\din\gg1$ case, the growth rate and the shear across $\din$ are comparable, and there is no small parameter with which to expand the equations. The current work is therefore less rigorous and based essentially on a scaling argument, rather than an analytic solution. Nevertheless, we show that our argument correctly describes the numerical simulations.

\section{Basic equations}
For our analysis, we use the reduced MHD equations in two dimensions,
\begin{align}
\frac{\partial\nabla_\perp^2\Phi}{\partial t} + \left\{\Phi,\nabla^2\Phi\right\} - \{\Psi,\nabla_\perp^2\Psi\} = 0,\label{eq:Phirmhd}\\
\frac{\partial \Psi}{\partial t} + \{\Phi,\Psi\} -\eta\nabla_\perp^2\Psi= 0,\label{eq:Psirmhd}
\end{align}
where the Poisson bracket $\{f,g\}=\hat{\boldsymbol{z}} \cdot(\nabla_\perp f \times \nabla_\perp g)$. From the flux and stream functions $\Psi$ and $\Phi$, we obtain the perpendicular (to $\hat{\vz})$ magnetic field (in velocity units) $\vb_\perp = \hat{\vz}\times \nabla_\perp \Psi$ and the perpendicular velocity $\vu_\perp = \hat{\vz}\times \nabla_\perp \Phi$. We linearize the equations around the equilibrium
\begin{equation}
\Phi_0 = \alpha \Psi_0, \quad b_{0y} = \vAy f(x/a) = \partial_x \Psi_0,
\end{equation}
which is the same as mentioned previously (Eqs.~\ref{eq:eqb} and \ref{eq:equ}), and assume fluctuations of the form 
\begin{align}
\delta \Phi &= \Phi(x) \exp(iky + \gamma t),\\
\delta \Psi &= \Psi(x) \exp(iky + \gamma t),
\end{align}
obtaining
\begin{align}
\left(\gamma +i\alpha k\vAy f\right)\left(\Phi'' - k^2\Phi\right) - i\alpha k\vAy f''\Phi - ik\vAy f\left(\Psi''-k^2\Psi- \frac{f''}{f}\Psi\right)&=0,\label{eq:linearphi}\\
\left(\gamma +i\alpha k\vAy f\right)\Psi -ik\vAy f \Phi - \eta\Psi''     &=0,\label{eq:linearpsi}
\end{align}
where we have assumed $ka\ll1$ and $a\gg \delta_{in}$ and thus we may neglect terms proportional to $\eta k^2$, and we denote x-derivatives as $\partial_x A = A'$.
\section{Outer region}
Far from the layer, where $x\sim a \gg \delta_{in}$, we may neglect all the terms involving the resistivity $\eta$. If we also assume $\gamma \ll k\vAy$ (which will be confirmed later), we may also neglect the growth terms, and obtain
\begin{align}
\alpha\left(\Phi'' - k^2\Phi - \frac{f''}{f}\Phi\right)&=\left(\Psi''-k^2\Psi- \frac{f''}{f}\Psi\right).\\
\alpha\Psi &= \Phi,\label{eq:outerphi}
\end{align}
Inserting the latter in the former, we obtain
\begin{equation}
(1-\alpha^2) f [\Psi'' - k^2\Psi - (f''/f)\Psi]=0,\label{eq:outer}
\end{equation}
which is just the constant $1-\alpha^2$ times the equivalent equation for the tearing mode with no equilibrium flow shear: thus, the solution for $\Psi$ is unchanged in the outer region. The solution for $\Phi$, however, is very different, and is given by Eq.~\ref{eq:outerphi}.

As $x\to 0$, $\partial_x \gg k$ and $f\approx x/a$, and of the outer equation all we are left with is
\beq
\Psi''=0,
\eeq
whence 
\beq
\Psi \to \Psi_\infty(1+ \frac{1}{2}\Delta'|x|), \quad x\to 0,
\eeq
defining
\beq
\Delta' = \frac{[\Psi']_{-0}^{+0}}{\Psi(0)},\label{eq:deltaprime}
\eeq
the discontinuity in the outer solution's magnetic field across the inner region. 

\section{Inner region}
In the inner region, of width $\delta_{in}\ll a$ (which will be determined later), the microphysical terms become important. Here, 
\begin{equation}
    x\ll a, f\approx x/a,\quad \text{and}\,\, \partial^2/\partial x^2 \gg k^2, f''/f.
\end{equation}
Defining
\begin{equation}
    \delta = \frac{\gamma a}{k\vAy}, \delta_\eta = \left(\frac{\eta a}{k \vAy}\right)^{1/3},
\end{equation}
we obtain the inner region equations
\begin{align}
    %(i\delta -\alpha x)\Phi'' &= - x\Psi'',\\
    (\delta + i\alpha x)\Phi''-ix\Psi'' &= 0,\label{eq:innerphi}\\
    %(i\delta - \alpha x)(\Psi-d_e^2\Psi'') &= x(\rho_s^2\Phi''-\Phi).\\
    (\delta + i\alpha x)\Psi -ix\Phi &= \delta_\eta^3\Psi''.\label{eq:innerpsi}
\end{align}
To match the inner solution with the outer solution, we integrate (\ref{eq:innerphi}) to obtain
\begin{equation}
\delta\int_{-\infty}^\infty \frac{\Phi''}{x} dx =  i\Psi'|_-^+-i\alpha\Phi'|_-^+,
\end{equation}
where the integral is to be understood to be over the inner solution from $x\ll-\delta_{in}$ to $x\gg\delta_{in}$, and the jump in a quantity over the inner layer is denoted $|_-^+$. To evaluate the RHS, we can use the asymptotic behaviour of the velocity in the outer region, (\ref{eq:outerphi}), as well as the definition of $\Delta'$ (\ref{eq:deltaprime}), to obtain
\begin{equation}
    \delta\int_{-\infty}^\infty \frac{\Phi''}{x} dx = i\left(1-\alpha^2\right)\Delta'\Psi_\infty.\label{eq:match}
\end{equation}
We now normalize the equations by the (as yet undetermined) $\delta_{in}$, writing
\begin{equation}
    \xi=\frac{x}{\delta_{in}},\quad\lambda = \frac{\delta}{\delta_{in}},
\end{equation}
and denoting $\partial_\xi A = A'$. We obtain
\begin{align}
    (\lambda + i\alpha\xi) \Phi''-i\xi\Psi'' &= 0,\label{eq:innernormphi}\\
    (\lambda + i \alpha \xi)\Psi - i\xi\Phi &= \frac{\delta_\eta^3}{\delta_{in}^3}\Psi''.\label{eq:innernormpsi1}
\end{align}
In the rescaled variables, the matching condition (\ref{eq:match}) reads
\begin{equation}
    \lambda\int_{-\infty}^\infty \frac{\Phi''}{\xi} d\xi = i\left(1-\alpha^2\right)\Delta'\delta_{in}\Psi_\infty.\label{eq:matchnorm}
\end{equation}
Substituting (\ref{eq:innernormphi}) into (\ref{eq:innernormpsi1}), we have
\begin{equation}
    (\lambda + i \alpha \xi)\Psi - i\xi\Phi = - i 
    \frac{\delta_\eta^3}{\delta_{in}^3}(\lambda+i\alpha\xi)\frac{\Phi''}{\xi}.\label{eq:innernormpsi2}
\end{equation}
Dividing by $\xi$, differentiating twice, and then using (\ref{eq:innernormphi}) again and dividing by $1-\alpha^2$,
\begin{align}
    \frac{2\lambda}{1-\alpha^2} \frac{\Psi-\xi\Psi'}{\xi^3} &+\left(\frac{2\lambda\alpha}{1-\alpha^2}-\frac{i\lambda^2}{\xi(1-\alpha^2)}-i\xi\right)\frac{\Phi''}{\xi}%\nonumber\\
    %&
    = -\frac{i}{1-\alpha^2} \frac{\delta_\eta^3}{\delta_{in}^3}\left(
    (\lambda+i\alpha\xi)\frac{\Phi''}{\xi^2}\right)''.\label{eq:innermaster}
\end{align}
For the shear flow to amount to more than a small correction to the growth rates, we require the growth rate to be small compared to the shear rate across the inner layer, $\gamma \ll \alpha k \vAy \din /a$. This assumption may be conveniently written as
\begin{equation}
    \lambda\ll\alpha.\label{eq:shearcomp}
\end{equation}
The tearing mode without shear has $\lambda \ll1$ for $\Delta'\din\ll1$ \citep{fkr}, but $\lambda\sim1$ for $\Delta'\din\ll1$ \citep{coppi}. \cite{boldyrev2018} note that if $\lambda \gg \alpha$, the shear flow cannot affect the structure of the inner layer, and conclude that the growth rate should be unchanged for $\Delta'\din\gg1$. However, given that at most $\lambda \sim 1$, it is worth reexamining the behaviour for $\alpha$ close to unity.

\section{Scaling theory}
We can now choose the inner lengthscale as
\begin{equation}
    \delta_{in} = \delta_\eta\left(\frac{\alpha}{1-\alpha^2}\right)^{1/3} = \left(\frac{\alpha\eta a}{(1-\alpha^2)k\vAy}\right)^{1/3}.\label{eq:deltainres}
\end{equation}
Inserting this choice and dividing by $\alpha$ makes the coefficient on the RHS of (\ref{eq:innermaster}) unity,
\begin{align}
    \frac{2\lambda}{1-\alpha^2} \frac{\Psi-\xi\Psi'}{\xi^3} &+\left(\frac{2\lambda\alpha}{1-\alpha^2}-\frac{i\lambda^2}{\xi(1-\alpha^2)}-i\xi\right)\frac{\Phi''}{\xi} = -i\left(\left[\frac{\lambda}{\alpha\xi}+i\right]\frac{\Phi''}{\xi}\right)''.
\end{align}
By comparing terms, we obtain the scaling for $\xi\gg1$
\begin{equation}
    \frac{\Phi''}{\xi}\sim \frac{i\lambda}{1-\alpha^2}\frac{\Psi}{\xi^4},
\end{equation}
where we have assumed not only $\lambda\ll\alpha$, but also that $\lambda/(1-\alpha^2)$ is at most order unity. We will confirm this later. To match to the outer solution using Eq.~(\ref{eq:matchnorm}), we need an estimate for the size of $\Psi$ inside the inner layer. The outer solution at $x=\pm\delta_{in}$ is
\begin{equation}
    \Psi_{\rm{outer}}(\delta_{in}) = \Psi_\infty\left(1+\frac{1}{2}\Delta'\delta_{in}\right),
\end{equation}
and the resistive terms ``smooth" the singularity on this scale: thus, we estimate
\begin{equation}
    \Psi\sim \begin{cases}
        \Psi_\infty, \quad&\Delta'\delta_{in}\ll1,\\
        \Delta'\delta_{in}\Psi_\infty, \quad&\Delta'\delta_{in}\gg1.
    \end{cases}
\end{equation}
The former assumption is the famous "constant-$\Psi$" approximation of \cite{fkr}. The matching integral is over an interval in $\xi$ of order unity, and so 
\begin{equation}
    \lambda\int_{-\infty}^\infty \frac{\Phi''}{\xi} d\xi \sim \frac{i \lambda^2}{1-\alpha^2}\Psi.
\end{equation}
Inserting our estimate for $\Psi$ and comparing to the matching condition (\ref{eq:matchnorm}), we obtain
\begin{equation}
    \lambda^2 \sim \begin{cases}
        \Delta'\delta_{in} (1-\alpha^2)^2, \quad & \Delta'\delta_{in}\ll1 \\
        (1-\alpha^2)^2, \quad & \Delta'\delta_{in}\gg1.
    \end{cases}\label{eq:lambda}
\end{equation}
Note that the two scalings match at $\Delta'\delta_{in}\sim 1$. Since we have assumed that $\lambda\ll\alpha$, the latter scaling is only valid for
\begin{equation}
    1-\alpha^2\ll\alpha,
\end{equation}
or $\alpha$ quite close to $1$ (see Sec.~\ref{sec:smallshear}). Inserting (\ref{eq:deltainres}), in terms of $\delta$,
\begin{equation}
    \delta \sim \begin{cases}
        \Delta'^{1/2}\alpha^{1/2}(1-\alpha^2)^{1/2}\delta_\eta^{3/2}, \quad & \Delta'\delta_{in}\ll1, \\
        \alpha^{1/3}(1-\alpha^2)^{2/3}\delta_\eta, \quad & \Delta'\delta_{in}\gg1,
    \end{cases}
\end{equation}
or, in terms of more physical variables,
\begin{equation}
    \frac{\gamma a}{\vAy} \sim \begin{cases}
        \alpha^{1/2}(1-\alpha^2)^{1/2}(\Delta'a)^{1/2}(ka)^{1/2}S^{-1/2}, \quad & \Delta'\delta_{in}\ll1, \\
        \alpha^{1/3}(1-\alpha^2)^{2/3}(ka)^{2/3}S^{-1/3}, \quad & \Delta'\delta_{in}\gg1,\label{eq:sheargamma}
    \end{cases}
\end{equation}
The scaling for $\Delta'\delta_{in}\ll1$ is the same as the growth rate obtained by \cite{chen1989}, up to a prefactor of order unity. The dependence of the scaling on $\alpha$ for $\Delta'\delta_{in}\gg1$ is new: \cite{chen1989} and \cite{boldyrev2018} both conclude (correctly) that once $\lambda\sim 1$, the scaling with $S$ should be unaffected by flow shear: we have shown here that the growth rate is suppressed as $1-\alpha^2\to0$.

For $\Delta'\din \ll 1$, the growth rate depends on the $k$-dependence of $\Delta'a$, with $\Delta'a\propto (ka)^{-n}$ for $ka\ll1$, a property of the equilibrium profiles:
\begin{equation}
    \Delta'\din \ll1 : \quad \frac{\gamma a }{\vAy} \sim \begin{cases}
\alpha^{1/2}(1-\alpha^2)^{1/2}S^{-1/2}, \quad &n=1,\\
\alpha^{1/2}(1-\alpha^2)^{1/2}S^{-1/2}(ka)^{-1/2}, \quad &n=2.
    \end{cases}
\end{equation}
Equating the two growth rate expressions for $\Delta'\din\ll1$ and $\Delta'\din\gg1$, the transition between the two  occurs at a wavevector
\begin{align}
    k_{tr} a \sim \begin{cases}
        \left(\frac{\alpha}{1-\alpha^2}\right)^{1/4}S^{-1/4}, \quad & n=1,\\
        \left(\frac{\alpha}{1-\alpha^2}\right)^{1/7}S^{-1/7},\quad & n=2,
    \end{cases}\label{eq:ktr}
\end{align}
For $1-\alpha^2\ll1$, the transitional wavenumber increases with $\alpha$. This is the opposite behavior predicted by \cite{chen1989}, who predicted that the tearing mode stabilized as $\alpha\to1$ via the constant-$\Psi$ mode (\ref{eq:sheargamma} for $\Delta'\din\ll1$) becoming valid at increasingly small $k$. In our analysis, because the $\Delta'\din\gg1$ growth rate also depends on $\alpha$, $k_{tr}$ increases with $\alpha$ and the nonconstant-$\Psi$ tearing mode becomes valid at progressively larger $k$. At the transitional wavevector, the growth rate is
\begin{align}
    \frac{\gamma_{tr}a}{\vAy} \sim \begin{cases}
        \alpha^{1/2}(1-\alpha^2)^{1/2}S^{-1/2}, \quad & n=1,\\
        \alpha^{3/7}(1-\alpha^2)^{4/7} S^{-3/7}, \quad & n=2.
    \end{cases}\label{eq:maxgrowth}
\end{align}
This is also the maximum growth rate of the instability. Finally, the width of the inner layer is given by Eq.~(\ref{eq:deltainres}),
    \begin{equation}
        \frac{\delta_{in}}{a} \sim \left(\frac{\alpha}{1-\alpha^2}\right)^{1/3} (ka)^{-1/3} S^{-1/3},
    \end{equation}
    irrespective of $\Delta'$.

\begin{figure}
    \centering
    \includegraphics[width=0.8\linewidth]{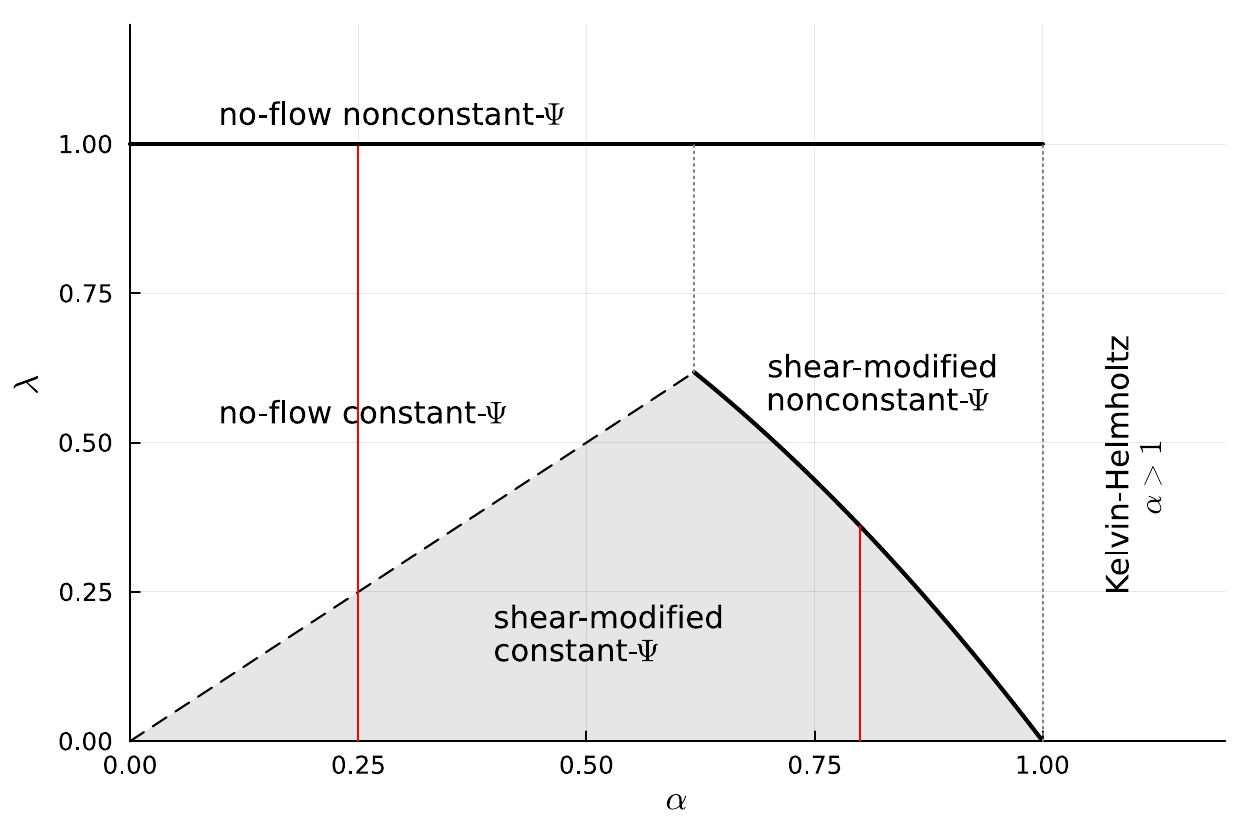}
    \caption{Different parameter regimes in the $\lambda$-$\alpha$ plane. As $k$ decreases from a large value, $\lambda$ initially increases, until it hits one of the thick black lines corresponding to the nonconstant-$\Psi$ scalings for $\Delta'\din\gg1$ in Eq.~\ref{eq:lambda}. For large enough $k$ (small enough $\lambda$), the mode always starts in the shear-modified constant-$\Psi$ regime (shaded region), provided $\alpha\neq 0$. For relatively small $\alpha$ (e.g. the red line at $\alpha=0.25$), the assumption $\lambda\ll\alpha$ is violated along the dotted line, and the no-flow constant-$\Psi$ scalings become applicable, with the growth rate achieving a maximum at the transition into the no-flow nonconstant-$\Psi$ scalings once $\lambda\sim 1$. For larger $\alpha$ (e.g. the red line along $\alpha=0.8$), the maximum growth rate is attained in the shear-modified regime, with the shear-modified nonconstant-$\Psi$ scalings along the curved thick line. As $\alpha\to1$ from below, the mode is completely stabilized}
    \label{fig:diagram}
\end{figure}
\section{Transition to the ``no-flow" scalings}\label{sec:smallshear}

For the shear to be important and able to affect the scalings, the shear across the inner layer must be comparable to or greater than the growth rate, cf. Eq.~(\ref{eq:shearcomp}).
At and below the transitional wavenumber between the small- and large-$\Delta'$ scalings, $\lambda \sim 1-\alpha^2$: thus, for the shear to be still important there, we must have
\begin{equation}
    1-\alpha^2 \lesssim C\alpha,
\end{equation}
where $C$ is some constant that we have not determined. A diagram representing the $\lambda$-$\alpha$ plane is shown in Figure~\ref{fig:diagram}, where we have chosen $C=1$ for illustrative purposes. For very small $\lambda$, even a small shear is significant and the shear-modified constant-$\Psi$ scalings (\ref{eq:sheargamma}, $\Delta'\din\ll1$) apply. $\Delta'\din$ is a decreasing function of $k$. Using (\ref{eq:lambda}), as $k$ decreases, initially $\lambda$ increases moving vertically upwards on Figure~\ref{fig:diagram}. Once the transitional wavenumber (\ref{eq:ktr}) is reached, $\lambda$ remains constant as the nonconstant-$\Psi$ regime has been reached.

For relatively small $\alpha$, e.g. the red line at $\alpha=0.25$, at some point $\lambda>\alpha$ and there is a transition across the dashed line into the "no-flow" constant-$\Psi$ instability \citep{fkr}, with the prefactor of the scaling possibly modified slightly by the small shear \citep{chen1989}. $\lambda$ then increases until $\lambda\sim1$ along the thick line, where the mode becomes the no-flow nonconstant-$\Psi$ mode \citep{coppi}. This is the situation shown in Figure 5 of \cite{boldyrev2018}: the maximum growth rate is essentially unaffected by the shear flow.

However, for relatively large $\alpha$, e.g. the red line at $\alpha=0.8$, the mode stays in the shear-modified constant-$\Psi$ mode until $\lambda\sim1-\alpha^2$, at which point the mode obeys the shear-modified nonconstant-$\Psi$ scalings, (\ref{eq:sheargamma}, $\Delta'\din\gg1$). In this case, the maximum of the growth rate \emph{is} affected by the shear, and is given by (\ref{eq:maxgrowth}). Thus, the growth rate (correctly) vanishes as $\alpha\to 1$. We stress that we are not predicting the exact location of the boundary between these two behaviours, which depends on the undetermined prefactors in the growth rates and in the exact expressions for $\Delta'$. It is safest to assume that the maximum growth rate only occurs in the shear-modified regime when $1-\alpha^2\ll1$.

Finally, we also mention that for $\alpha>1$, there is no tearing mode, but instead the Kelvin-Helmholtz instability is active: the growth rate of this instability is extremely fast compared to the tearing mode.

\begin{figure}
    \centering
    \includegraphics[width=0.8\linewidth]{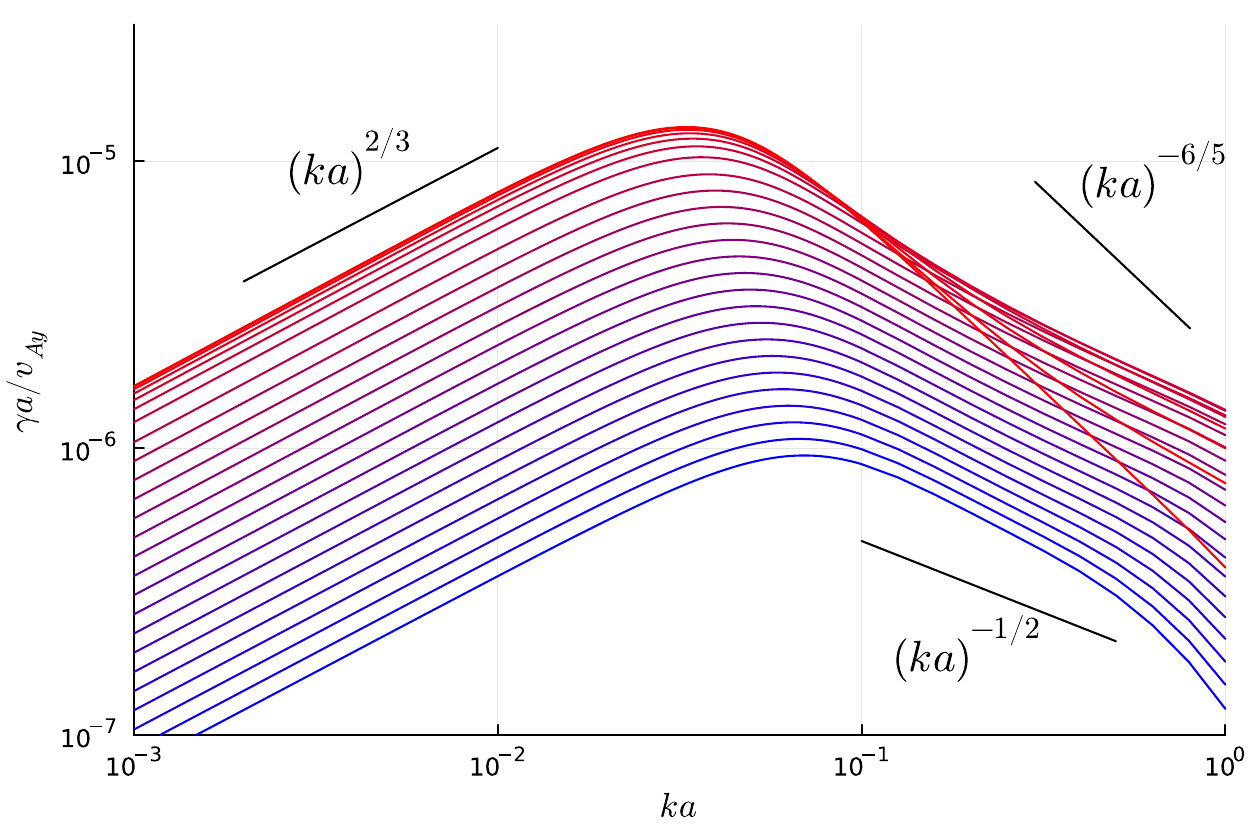}
    \caption{Growth rate of the tearing mode, $\gamma$ versus wavenumber $k$ for different $\alpha$, from $\alpha=0$ (red) to $\alpha=\sqrt{1-0.01}=0.995$ (blue). Also marked are power laws in $k$ corresponding to the nonconstant-$\Psi$ tearing mode $(ka)^{2/3}$, the no-flow constant-$\Psi$ tearing mode $(ka)^{-{6/5}}$, and the shear-modified constant-$\Psi$ tearing mode $(ka)^{-1/2}$. For these data, $S=10^{12}$.}
    \label{fig:res_gammavsk_alphas}
\end{figure}
\section{Numerical tests}
We have written an eigenvalue code to solve Eqs.~(\ref{eq:linearphi}--\ref{eq:linearpsi}). To test the predictions above, we set $a=1$, $\vAy=1$, and vary $S=\eta^{-1}$, $k$, and $\alpha$. We use the profile
\begin{align}
    f(x) &= -2\tanh(x)\sech^2(x),
\end{align}
For $ka\ll1$, $\Delta'a \sim 15/(ka)^2$; i.e. $n=2$. This is useful for testing because in this case, the maximum growth rate is attained only at the transitional wavenumber, rather than for all $k>k_{tr}$ as would be the case for the more usual $f(x)=\tanh(x)$ profile, for which $n=1$.

Figure \ref{fig:res_gammavsk_alphas} shows the growth rate as a function of $k$ for different $\alpha$. In detail, the curves shown correspond to 
\begin{align}
    \alpha = 0, 0.1, 0.2, \ldots 0.6, \quad \alpha=\sqrt{1-q},\\
    q = 10^{-3/10}, 10^{-4/10},\ldots 10^{-2},
\end{align}
with the small-$\alpha$ end spaced evenly in $\alpha$ and the large-$\alpha$ end spaced logarithmically in $1-\alpha^2$, up to $1-\alpha^2=0.01$. $\alpha$ increases from $\alpha=0$ (reddest) to $\alpha=\sqrt{1-0.01}=0.995$ (bluest). As can be seen by comparing to the power laws marked, the predicted scalings with $k$ agree quite well with the numerical solution. For small $\alpha$, the growth rate obeys the no-flow scalings,(\ref{eq:resnoflow})\citep{fkr,coppi}, while at larger $\alpha$, at small $k$ the growth rate transitions towards the shear-modified constant-$\Psi$ scaling (\ref{eq:sheargamma} for $\Delta'\din\ll1$). For $1-\alpha^2\ll1$, the growth rates also decrease with increasing $\alpha$, as predicted, and the wavenumber corresponding to the maximum growth rate moves to larger $k$, as predicted by (\ref{eq:ktr}).

\begin{figure}
    \centering
    \begin{minipage}{\linewidth}
    \includegraphics[width=0.45\linewidth]{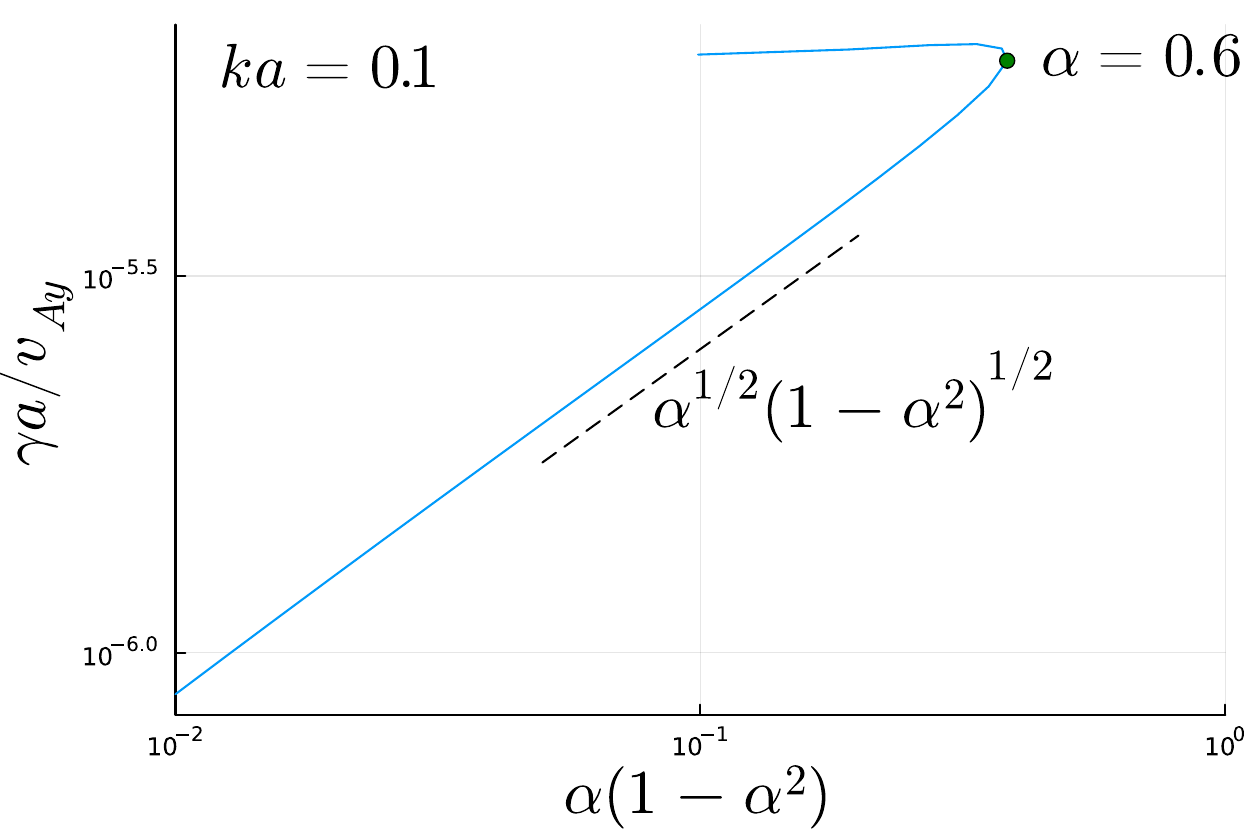}  \includegraphics[width=0.45\linewidth]{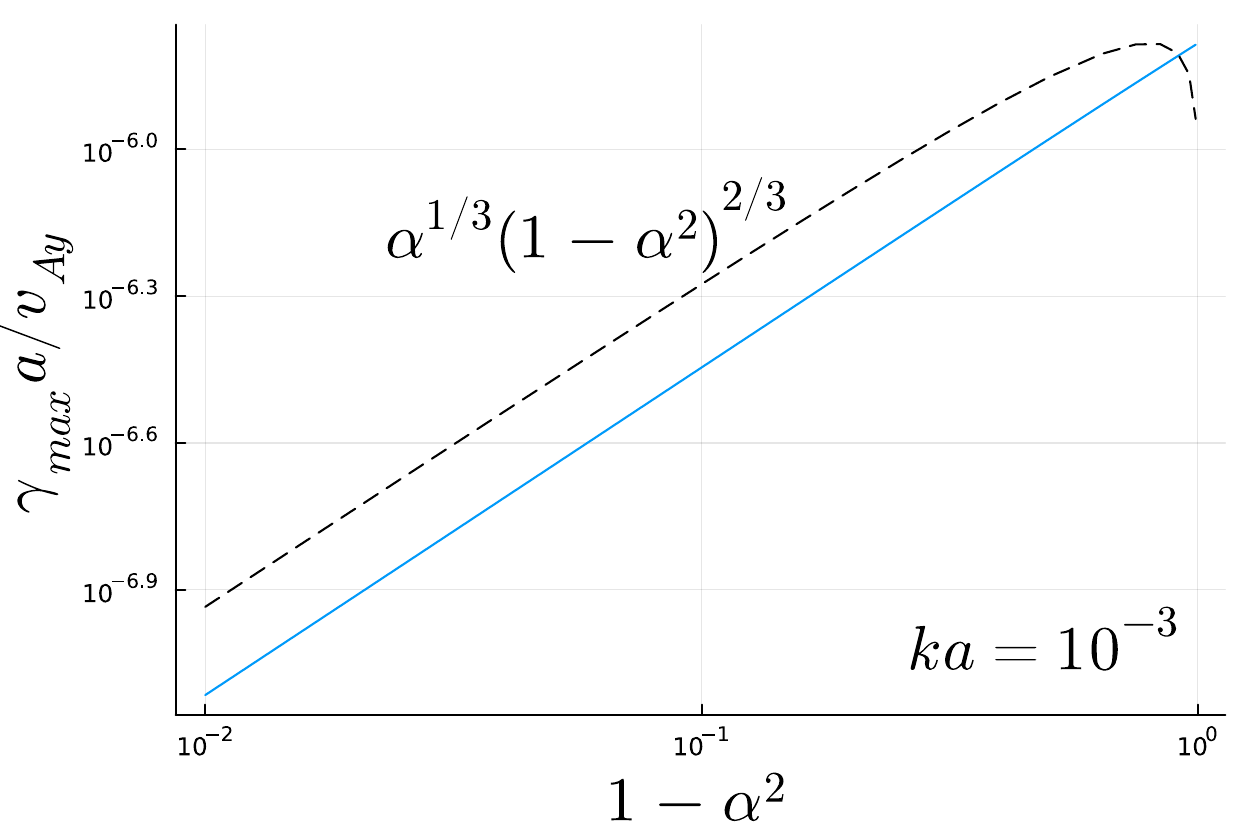} \\
    \includegraphics[width=0.45\linewidth]{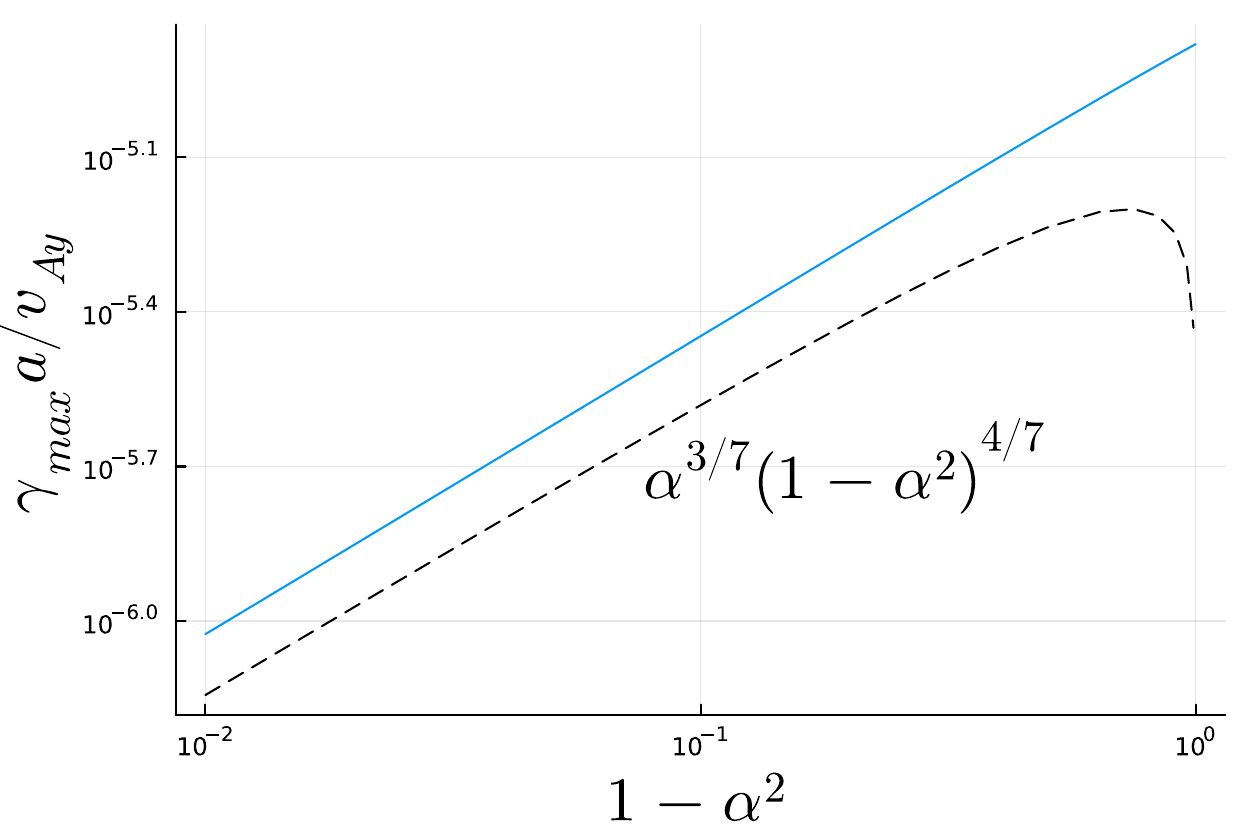} \includegraphics[width=0.45\linewidth]{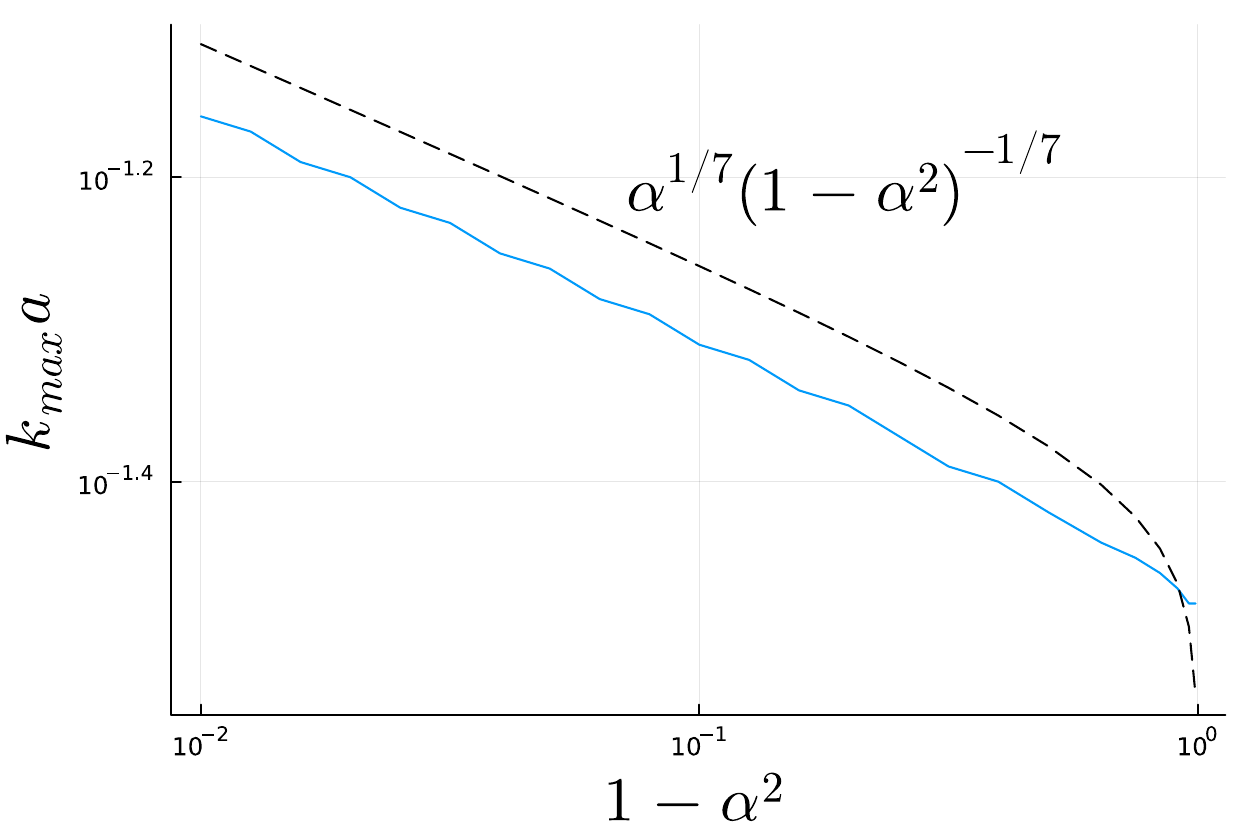}
    \end{minipage}
    \caption{Top left: $\gamma$ for $ka = 0.1$, in the constant-$\Psi$ region. Top right: $\gamma$ for $ka=10^{-3}$, in the nonconstant-$\Psi$ region. Bottom left: maximum growth rate $\gamma_{max}$. Bottom right: the wavenumber $k_{max}$ at which $\gamma_{max}$ is attained.}
    \label{fig:alphascalings}
\end{figure}

To test our predictions for how the growth rate scales with $\alpha$, we have plotted growth rates at fixed $k$ versus $\alpha$ in Figure \ref{fig:alphascalings}. For sufficiently large $\alpha\gtrsim 0.6$, the scalings agree very well with the predictions.

\begin{figure}
    \centering
    \begin{minipage}{\linewidth} 
    \includegraphics[width=0.45\linewidth]{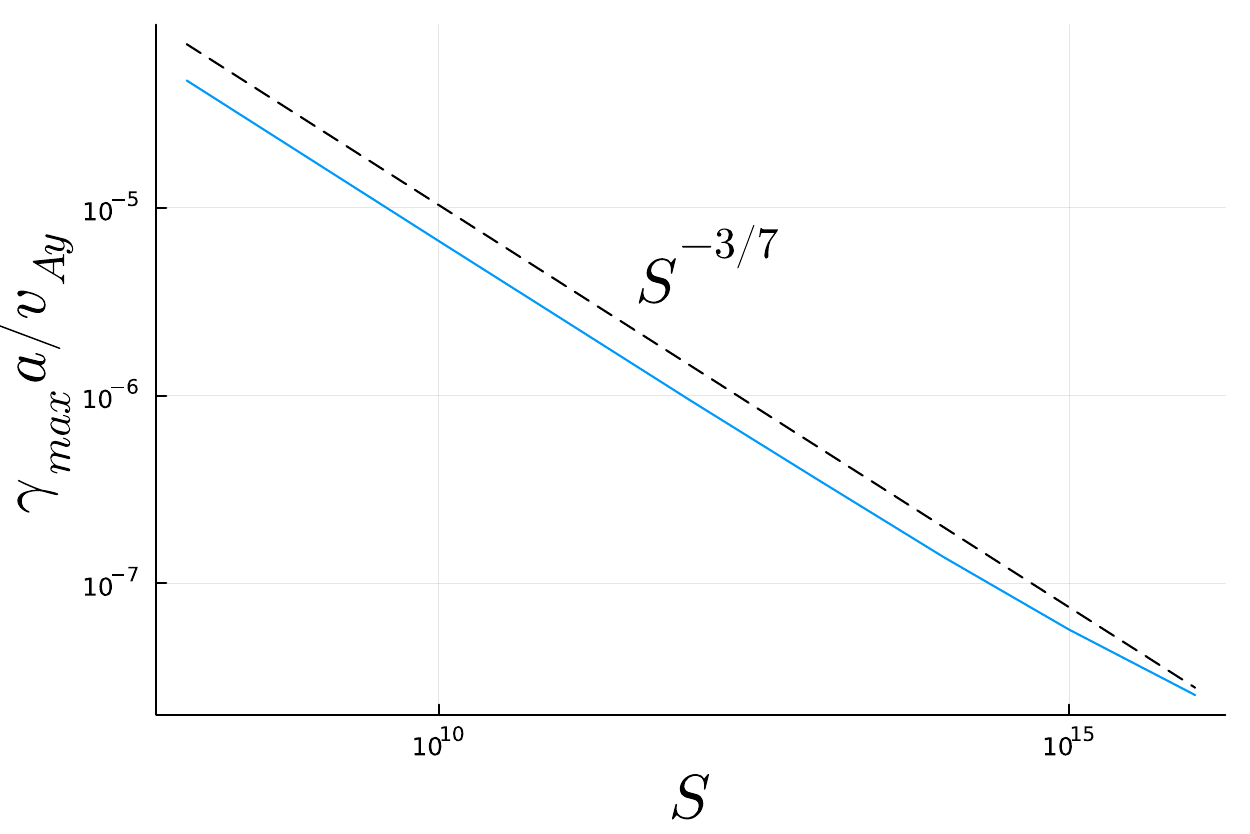}
    \includegraphics[width=0.45\linewidth]{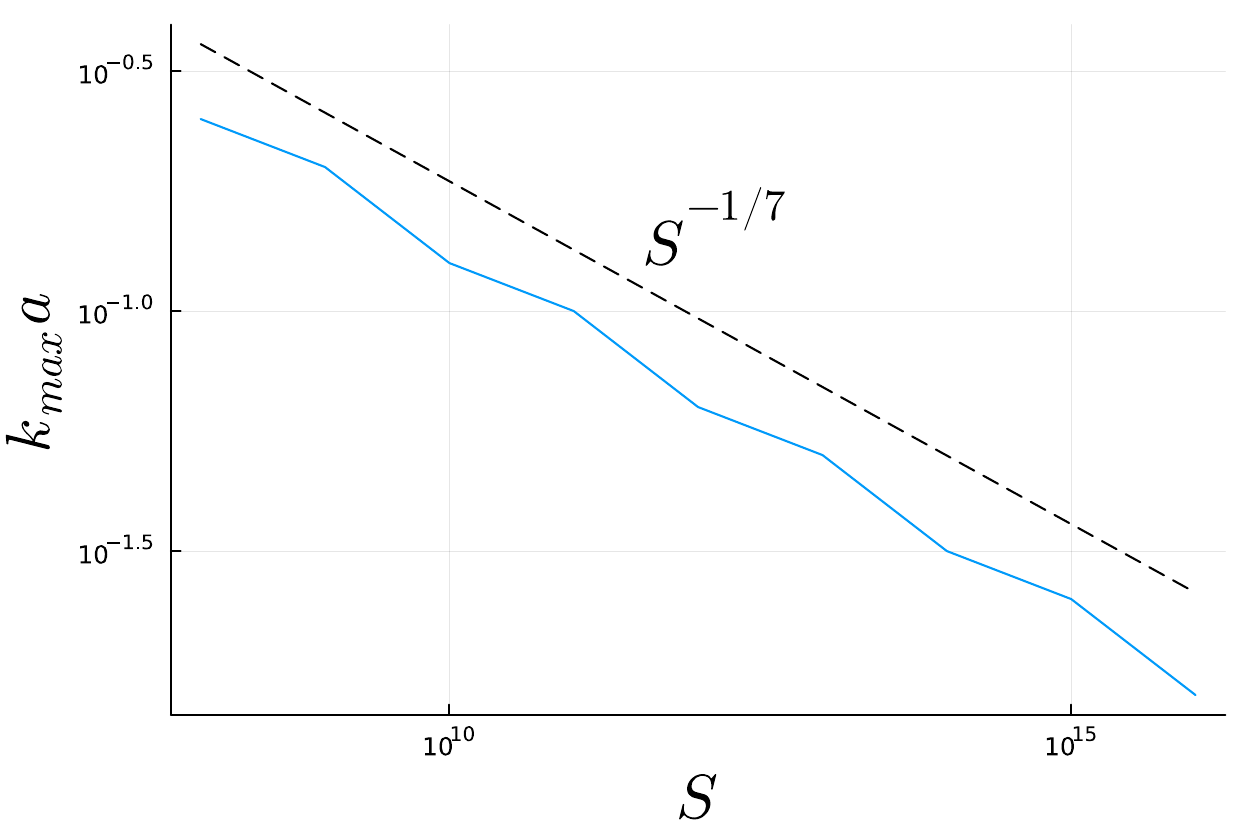}
    \end{minipage}
    \caption{Tearing mode scalings for $1-\alpha^2=0.01$. Left: $\gamma_{max}$ as a function of $S$. Right: $k_{max}$ as a function of $S$.}
    \label{fig:Sscalings}
\end{figure}

We also check the dependence of $\gamma$ on $S$ for large shear, $1-\alpha^2=0.01$, by scanning from $S=10^8$ to $S=10^{16}$. Figure \ref{fig:Sscalings} shows the expected scaling of the maximum growth rate $\gamma_{max}\propto S^{-3/7}$ and also the expected scaling of $k_{max}\propto S^{-1/7}$, unchanged from the no-flow case.

\begin{figure}
    \centering
    \begin{minipage}{\linewidth} 
    \includegraphics[width=0.45\linewidth]{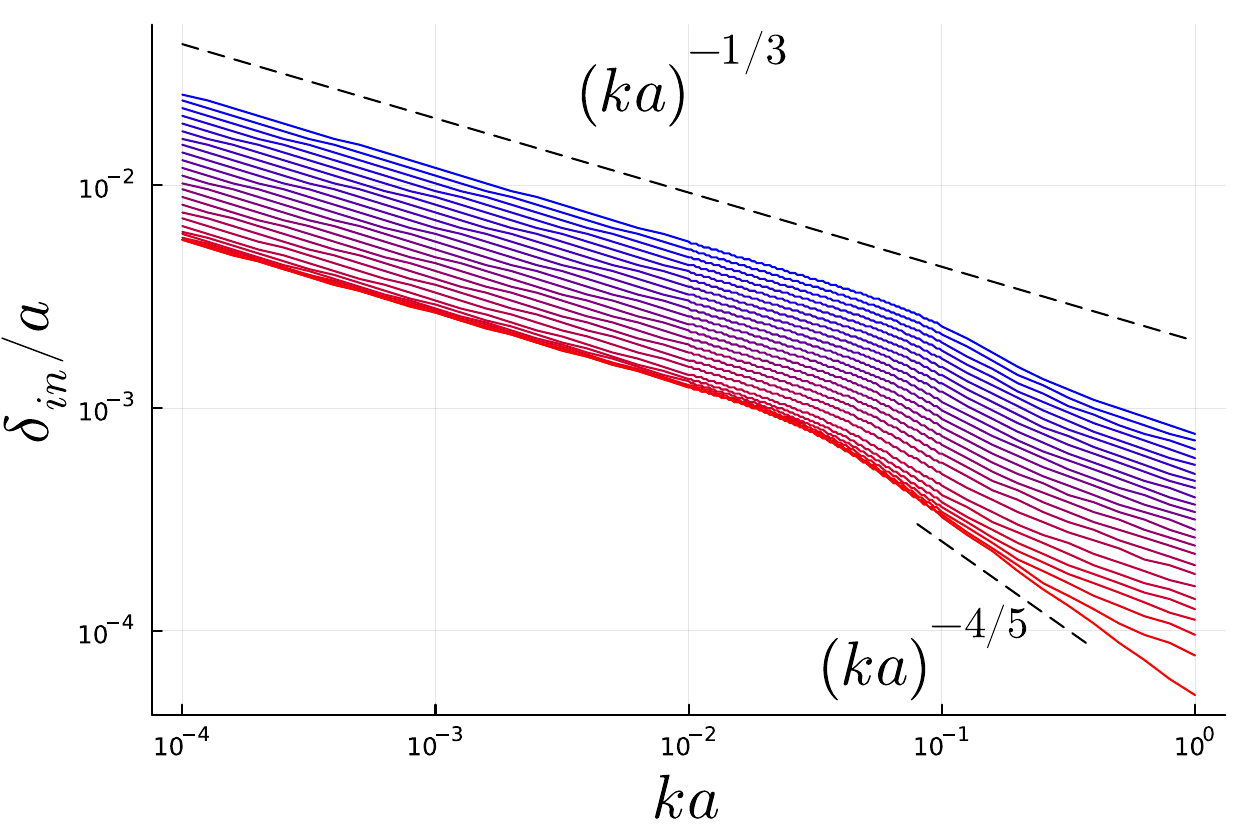}
    \includegraphics[width=0.45\linewidth]{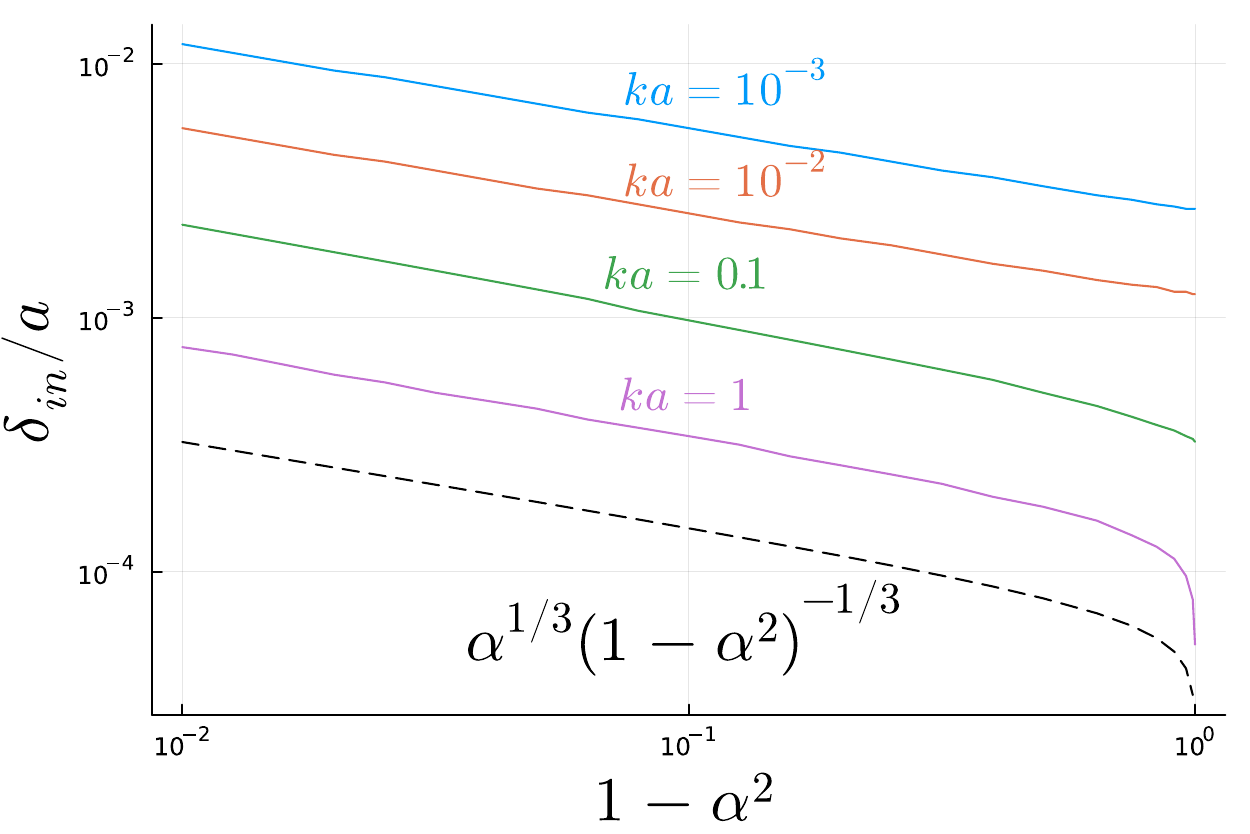}
    \end{minipage}
    \caption{The scaling of $\delta_{in}$ as a function of $k$ (left) and $\alpha$ (right). On the left, the colors represent $\alpha$, with the smallest $\alpha=0$ in red and the largest $\alpha\approx0.995$ in blue: the same color scale as in Figure~\ref{fig:res_gammavsk_alphas}. On the right, the different lines correspond to different wavenumbers, with the dotted line the predicted scaling.}
    \label{fig:dinscalings}
\end{figure}

Finally, we have also checked the scaling of the inner region. We define $\delta_{in}$ numerically as the width at which $\Psi''$ drops to $1/4$ of its maximum value. The dependence on both $k$ and $\alpha$, shown in Fig.~\ref{fig:dinscalings}, agree quite well with the predicted scalings, with $\delta_{in}\propto k^{-1/3}$ for significant shear and/or large $\Delta'\delta_{in}$ and $\delta_{in}\propto k^{-4/5}$ for small flow shear and $\Delta'\delta_{in}\ll1$. The exception is around the maximum of the growth rate, which could be caused by a mismatch in numerical prefactors between the small- and large- $\Delta'$ scalings, which we cannot predict with this scaling theory.

\section{Conclusions}
In this work, we have developed a scaling theory for the resistive tearing mode with flow shear along the current sheet, $u_{0y} = \alpha b_{0y}$, where $0<|\alpha|<1$ (see Eq.~\ref{eq:equ}). Previous work by \cite{chen1989} only studied the effect of shear on the growth rate in the weakly unstable "constant-$\Psi$" case where $\Delta'\din\ll1$. We have extended this to the non-constant-$\Psi$ case where $\Delta'\din\gg1$, but where $1-\alpha^2\ll1$, showing that the instability in this regime is also affected by the flow shear. We are then able to estimate the maximum growth rate of the instability over all $k$,
\begin{align}
    \frac{\gamma_{tr}a}{\vAy} \sim \begin{cases}
        \alpha^{1/2}(1-\alpha^2)^{1/2}S^{-1/2}, \quad & n=1,\\
        \alpha^{3/7}(1-\alpha^2)^{4/7} S^{-3/7}, \quad & n=2.
    \end{cases}\label{eq:maxgrowth2}
\end{align}
where $n$ depends on the specific equilibrium profile. There are two main conclusions. First, the shear does not affect the scaling of the maximum growth rate with $S$ (compare with Eq.~\ref{eq:maxnoshear}, the maximum growth rate for $n=1$ with no shear), in agreement with earlier work \citep{chen1989,boldyrev2018}. Second, the growth rate of the tearing mode vanishes as $\alpha\to 1$: while this is also predicted by \citep{chen1989}, the way in which this occurs is different. The earlier work by \cite{chen1989} and \cite{boldyrev2018} assumed that for $\Delta'\din\gg1$ the shear could not affect the growth rate, and thus the scaling of the maximum growth rate with $\alpha$ was not correctly described. We have shown that for $\alpha$ sufficiently close to $1$, the growth rate for $\Delta'\din\gg1$ actually decreases as $\gamma \propto (1-\alpha^2)^{2/3}$, slightly more strongly suppressed than the scaling for $\Delta'\din\ll1$, $\gamma\propto (1-\alpha^2)^{1/2}$. This means that as $\alpha\to1$, the wavenumber corresponding to the maximum of the growth rate is an \emph{increasing} function of $\alpha$ (see Eq.~\ref{eq:ktr}), unlike in the earlier theory of \cite{chen1989}, who predicted the opposite. We have validated our scaling theory with numerical simulations, finding good agreement with all of our predictions.

What does our theory mean for turbulence theories that involve disruption of turbulent structures by the tearing instability at small scales \citep{msc_disruption,loureiroboldyrev}? Before commenting on this, it is worth pointing out that resistive MHD is inapplicable to many situations of interest: one well-observed example of plasma turbulence in nature, the solar wind \citep{chen2016}, involves a plasma that is nearly collisionless. The tearing mode is then affected by the dispersive behaviour entering at the ion gyroradius, and reconnection is due not to resistivity but to electron-scale physics (e.g. electron inertia). We have examined the effects of flow shear on the collisionless tearing mode in \citep{mallet2025}, finding $\gamma_{tr}\propto 1-\alpha^2$. Collisional MHD may be appropriate for modelling reconnection in the higher-density regions of the low solar atmosphere \citep{wargnier2022,kahil2022}, in which case the shear modification of the resistive tearing growth rate described here could be important.%: however, the turbulence in this region may not be sufficiently imbalanced for the shear modifications to the tearing growth rate to be important \citep{chandran2009}.

Supposing that resistive MHD is an applicable model to our (natural or numerically simulated) plasma of interest, we have shown that the scaling of the growth rate with $S$ is unchanged by the shear. Then, provided that $\alpha$ is not too close to unity in the turbulent structures, there is little modification needed to theories of reconnecting turbulence \citep{msc_disruption,loureiroboldyrev}: the $\alpha$-dependence of the growth rate amounts to a coefficient of order unity. However, in imbalanced turbulence, where $\delta {z}^+ \gg \delta {z}^-$ (or vice versa), where $\delta \boldsymbol{z}^\pm = \delta \vu \pm \delta \vb$, we have $1-\alpha^2 \sim \delta {z}^-/\delta {z}^+\ll1$, so that $\alpha$ is very close to $1$. This could then change the scale at which onset of reconnection occurs by an appreciable factor. Very briefly, one expects tearing-mediated turbulence when $\gamma_{tr}\tau_{nl}\sim 1$. In the simplest model of imbalanced turbulence, the nonlinear cascade time is given by $\tau_{nl}^+\sim \delta z^-/L$, in which case $\gamma_{tr}\tau_{nl} \propto (1-\alpha^2)^{-1/2}$, and the imbalance makes the transition to tearing-mediated turbulence easier to achieve. However, there is currently no widely accepted theory of imbalanced MHD turbulence (see \cite{schekochihin2021} for a review): we do not know how the length $L$ and width $a$ of the turbulence-produced current sheets depend on each other, nor even if the simple $\tau_{nl}$ given here is the relevant timescale. Moreover, as mentioned previously, many examples of turbulence in nature involve collisionless plasmas for which MHD is inapplicable. Nevertheless, it would be interesting to look at the onset of reconnection in imbalanced MHD turbulence, especially given that in the balanced-turbulence case theoretical expectations \citep{msc_disruption,loureiroboldyrev} have been borne out by numerical simulations \citep{dong2022}. We therefore plan to investigate how the modification of the collisionless and collisional tearing modes with shear affect the turbulent cascade in future work.

\textbf{Declaration of Interests.--} The authors report no conflict of interest.

\textbf{Acknowledgements.--} AM, SE and MS were supported by NASA grant 80NSSC20K1284. JJ was funded by the U.S. Department of Energy under Contract No. DE-AC02-09CH1146 via an LDRD grant.
\bibliographystyle{jpp}
\bibliography{mainbib2024}
\end{document}